\documentclass[twocolumn,preprintnumbers,aps]{revtex4}
\usepackage{epsfig}
\begin{document}

\title{ Substrate effects on quasiparticles and excitons in graphene nanoflakes }
\author{ Weidong Sheng and Mengchao Sun }
\affiliation{ State Key Laboratory of Surface Physics and Department of Physics, Fudan 
University, Shanghai 200433, China }
\author{ Aiping Zhou }
\affiliation{ Department of Mathematics and Physics, Nanjing Institute of Technology, Nanjing 211167, China }
\author{ S. J. Xu }
\affiliation{ Department of Physics, The University of Hong Kong, Pokfulam Road, Hong Kong, China }

\begin{abstract}
The effects of substrate on electronic and optical properties of triangular and hexagonal 
graphene nanoflakes with armchair edges are investigated by using a configuration 
interaction approach beyond double excitation scheme. The quasiparticle correction to the 
energy gap and exciton binding energy are found to be dominated by the long-range Coulomb 
interactions and exhibit similar dependence on the dielectric constant of the substrate, 
which leads to a cancellation of their contributions to the optical gap. As a result, the 
optical gaps are shown to be insensitive to the dielectric environment and unexpectedly 
close to the single-particle gaps.
\end{abstract}
\maketitle

% \section{Introduction}

Graphene, an artificial material discovered recently \cite{Novoselov}, is a promising 
candidate in future microelectronic devices due to its extraordinary electronic \cite{Zhou} 
and optical properties \cite{Mueller}. Recently, many theoretical interests have been 
attracted to the study of substrate influence on the electronic structure, thermal 
conductivity and growth mechanisms in bulk graphene \cite{Li, Ciraci} and graphene 
nanoribbons \cite{Liang, Chen}. Experimentally, the effect of semi-insulating and metal 
substrates has been investigated by using ultraviolet and far-infrared photoelectron 
spectroscopy \cite{Wu, Kim}.

Although bulk graphene has almost zero band-gap, a finite gap can be opened and even 
engineered by quantum confinement effect in graphene nanoribbons and nanoflakes \cite{Son}. 
Electron-electron interactions would further modify this quasiparticle gap into the optical 
gap, which is commonly known as the excitonic effect \cite{Han, Nandkishore, Sabio, 
Paananen}. Many-body perturbation theory and configuration interaction method have been 
applied to calculate exciton binding energies in quasi-one-dimensional graphene nanoribbons 
\cite{Louie, Lu}, and excitonic absorption in triangular graphene quantum dots with zigzag 
edges \cite{Dutta, Hawrylak}. Undoubtedly, the study of quasiparticle and excitonic effects 
in these structures requires a proper treatment of the dielectric screening effect 
\cite{Jang} from various substrates like $\mbox{SiO}_2$ \cite{Ishigami}, diamond 
\cite{Avouris}, SiC \cite{Strupinski} or other semi-insulating materials.

At present, however, there have been very few attempts to investigate substrate effects on 
electronic structure and optical properties in graphene nanoflakes. In this letter, we will 
explore how various substrates affects quasiparticle self-energies, exciton binding energies 
and optical gaps in graphene nanoflakes. An interesting question that how sensitive the 
optical transitions are to the dielectric environment in nanographene structures, which is 
believed to have both fundamental and practical importance, will be answered.

\begin{figure}
  \includegraphics[width=0.40\textwidth]{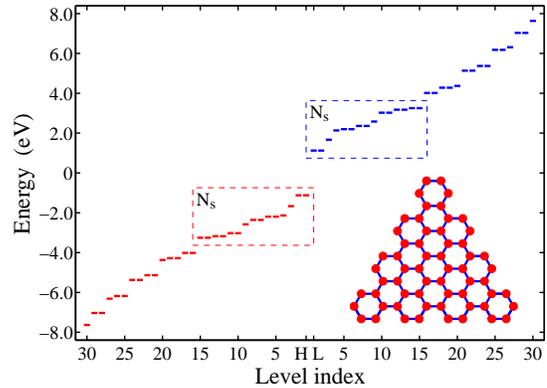}
  \caption{ Energy levels of a triangular graphene nanoflake (as shown in the inset). The
HOMO (highest occupied molecular orbital) and LUMO (lowest unoccupied molecular orbital)
states are denoted by $H$ and $L$, respectively. The number of electron (hole) states
(schematically shown in two dashed boxes) taken into account in the configuration
interaction computation is denoted by $N_s$. }
  \label{4e}
\end{figure}

% \section{Model and Method}

We consider two types of armchair graphene nanoflakes placed on various substrates such as 
$\mbox{SiO}_2$, diamond, and SiC. Figure \ref{4e} gives a schematic view of our first 
model system, a triangular graphene nanoflake. The number of carbon rings along each edge 
is set to be $N = 4$, which corresponds to a total number of atoms $n = 60$. The 
single-particle states are obtained by the use of the tight-binding model with the 
nearest-neighbor hopping. The matrix element of the single-particle Hamiltonian for 
electron $p$ is given by $ \langle i | \hat{\bf H}(p) | j \rangle = t$, if site $i$ and 
$j$ are the nearest neighbors, and would vanish if otherwise. The hopping energy is set to 
be -2.7~eV. The single-particle states, $\psi_m = \sum_{i=1}^{N} c_m^i | i \rangle$, are 
calculated by diagonalizing the Hamiltonian matrix and are plotted in Fig. \ref{4e}. A 
single-particle gap is seen to separate the occupied and unoccupied states. Moreover, the 
single-particle energies are found not continuous, instead, the energy levels form a 
series of clusters.

As electron-electron interactions exhibit different dimensional dependence in graphene 
from other semiconductors \cite{Kotov}, we make use of configuration interaction method to 
solve the interacting electron problem. Many-particle wave functions are expanded on the 
basis of single-particle states obtained previously by the tight-binding method. Unlike 
those structures with zigzag edges \cite{Hawrylak}, the nanoflake with armchair edges is 
seen to have a closed-shell energy spectrum with a well-defined energy gap. Therefore, we 
have to choose a number of valence states ($N_s$) from the HOMO (highest occupied 
molecular orbital) down and the same number of conduction states from LUMO (lowest 
unoccupied molecular orbital) up, as our basis to expand the following many-particle 
Hamiltonian,
\begin{eqnarray}
\hat{\bf H} &=& \sum_{p=1}^{N_e} \hat{\bf H}(p) + \frac{1}{\epsilon_r^*} \sum_{p \neq q}^{N_e} \hat{\bf V}(p,q) , \cr
\hat{\bf V}(p,q) &=& \frac{1}{4 \pi \epsilon_0} \cdot \frac{e^2}{\sqrt{|x_p-x_q|^2 + |y_p-y_q|^2}} ,
\end{eqnarray}
where $N_e$ is the number of electrons which equals to $N_s$ in our neutral half-filling 
system. As single-particle energy levels form a series of clusters, we choose the $N_s$-th 
level as the end of a cluster of states to ensure that $E_{N_s+1} - E_{N_s}$ is large 
enough compared with the pertinent Coulomb energies.

The effective background dielectric constant is determined by $\epsilon_r^* = 
\frac{1}{2}(\epsilon_r + 1)$ with $\epsilon_r$ for the substrate \cite{Jang, Siegel}. For 
$\mbox{SiO}_2$, diamond, and SiC, $\epsilon_r^*$ is given by $2.5$, $3.35$, and $6.4$, 
respectively. The Coulomb matrix elements \cite{Sheng} consist of the on-site and off-site 
parts as follows,
\begin{equation}
U_{pqrs} = \sum_{i=1}^N c_p^i c_q^i U_{00} c_r^i c_s^i +
\frac{1}{\epsilon_r^*} \sum_{i \neq j}^N c_p^i c_q^j \hat{\bf V}(i,j) c_r^i c_s^j ,
\label{cme}
\end{equation}
in which $U_{00} = 17.0$~eV is chosen for the on-site Coulomb interaction \cite{Wehling}. 
It is noted that only the off-site part is influenced by the dielectric screening. All the 
occupied states in the closed-shell system form a single reference configuration. For a 
given $N_s$, one can choose to move $m (\leq N_s)$ electrons from the occupied states to 
the unoccupied states, usually referred as a $m$-$th$ excitation, to construct a 
many-particle configuration. For the model systems considered in this work, we find that it is 
necessary to have $m \geq 5$ in order for the low-lying levels to be fully converged. Here 
we choose $(N_s,m) = (16,5)$ and the resulting sparse matrix has a dimension of 
$6,689,001$. ARPACK is used for the diagonalization of the matrix to obtain the energy 
levels $E_n(N_e)$ of the many-electron system.

For a given occupation number $N_e$, the quasiparticle gap can be then obtained by
\begin{equation}
E^{qp}_{gap} = \mu(N_e+1) - \mu(N_e) ,
\end{equation}
where $\mu(N_e)$ and $\mu(N_e+1)$ are the chemical potentials of the system defined by
\begin{eqnarray}
\mu(N_e) &=& E_0(N_e)-E_0(N_e-1) , \cr
\mu(N_e+1) &=& E_0(N_e+1)-E_0(N_e) ,
\end{eqnarray}
with $E_0(N_e)$ being the ground-state energy of the $N_e$-electron system. It is noted 
that the basis dimension of the system with either an extra electron ($N_e+1$) or hole 
($N_e-1$) increases by almost twice. The excitonic or optical gap \cite{Yang} is defined by

\begin{equation}
E^{op}_{gap} = E_1^{S=0}(N_e) - E_0^{S=0}(N_e) .
\end{equation}
The quasiparticle and optical gap is related by the exciton binding energy $E_X$ as follows,
\begin{equation}
E_X = E^{qp}_{gap} - E^{op}_{gap} .
\label{exciton}
\end{equation}

% \section{Result and Discussion}

\begin{table}
\caption{ List of quasiparticle gap and quasiparticle correction to the energy gap 
calculated for various substrates. }
\begin{ruledtabular}
\begin{tabular}{llcc}
Substrate       & $\epsilon_r^*$ & $E^{qp}_{gap}$ & $E^{qp}_{gap} - E^{sp}_{gap}$ \\
Silicon Carbide‎ & 6.4            &  3.2384        &  0.9930                       \\
Diamond         & 3.35           &  3.5911        &  1.3457                       \\
Silicon dioxide & 2.5            &  3.8257        &  1.5803                       \\
\end{tabular}
\end{ruledtabular}
\label{4q}
\end{table}

Table \ref{4q} lists the quasiparticle gap $E^{qp}_{gap}$ and quasiparticle correction to 
the energy gap $E^{qp}_{gap} - E^{sp}_{gap}$ calculated for various substrates. First we 
would like to mention that the chemical potential we calculate is only for the interacting 
electron system because the background ionic charges only shifts all addition energies in 
the same way and thus shall have little effect on the quasiparticle gap. Compared with the 
single-particle gap $E^{sp}_{gap} \approx 2.25$~eV, we see that the quasiparticle gaps are 
larger by about 0.99 to 1.58~eV due to strong electron-electron interactions. The Coulomb 
matrix elements averaged among the HOMO and LUMO states are found to be 1.47~eV (direct) and 
0.35~eV (exchange), which is either larger than or comparable with the corresponding kinetic 
energy $E^{sp}_{gap}/2 = 1.12$~eV. Moreover, a typical correlation element is found to be 
about one third of the exchange term and thus would make a non-negligible contribution to 
the total energy. As the substrate changes from SiC to $\mbox{SiO}_2$, we find that the 
quasiparticle self-energy correction to the energy gap, i.e., $E^{qp}_{gap} - E^{sp}_{gap}$, 
increases by about 60\% or from 0.99~eV to 1.58~eV. Considering that this increment occurs 
as a result of the reduction of the effective dielectric constant also by 60\%, we believe 
that the quasiparticle effect is dominated by the long-range Coulomb interaction which is 
controlled by $\epsilon_r^*$. Actually, if one removes the long-range Coulomb interaction by 
setting $\epsilon_r^* \rightarrow \infty$, $E^{qp}_{gap} - E^{sp}_{gap}$ would reduce to 
0.49~eV. In the case of $\mbox{SiO}_2$ substrate, this means that the on-site Coulomb 
interaction contributes only about 30\% of the overall quasiparticle effect.

\begin{figure}
  \includegraphics[width=0.40\textwidth]{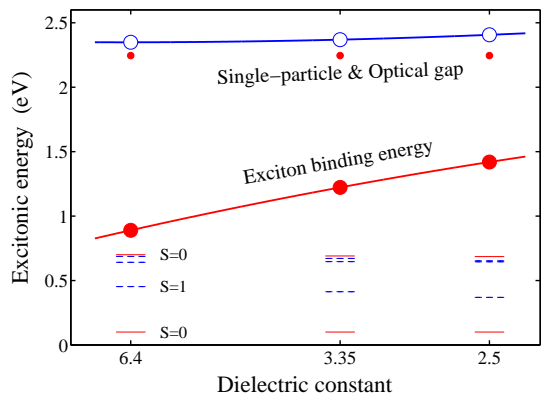}
  \caption{ Optical gap (in open dots) and exciton binding energy (in solid dots)
calculated as a function of the inverse of the effective dielectric constant. The
single-particle gap, which does not vary with the dielectric constant, is shown in the
smaller solid dots as a reference. Insets: Schematic view of the energy levels and their
total spins. Solid lines for spin singlets and dashed lines for spin triplets. }
  \label{4x}
\end{figure}

Figure \ref{4x} plots the energy spectra for the triangular model for three different 
substrates. Above the singlet ground state, we see three triplet ($S = 1$) states before 
the first excited state of $S = 0$. The calculated optical gap is plotted in open dots as 
a function of the effective dielectric constant. As a reference, the single-particle gap 
is shown in solid dots just below the optical gaps. When the substrate changes from SiC to 
$\mbox{SiO}_2$, the relative difference between $E^{op}_{gap}$ and $E^{sp}_{gap}$ is found 
to increase from 0.1~eV to 0.16~eV. However, this difference is so small that the optical 
gap is close to the single-particle gap and its absolute value increases by only 2.5\%. 
Compared with the quasiparticle gap, the substrate hence plays only a minor role in the 
optical gap. In other words, the optical gap is insensitive to the long-range Coulomb 
interactions. Then how about short-range interactions ? As the on-site Coulomb interaction 
$U_{00}$ reduces from 17.0~eV to 9.3~eV, we find that the optical gap decreases from 
2.41~eV to 2.39~eV by less than 1\%. Therefore, we can safely conclude that the optical 
gap is sensitive to neither the long-range nor short-range Coulomb interactions. In fact, 
we see that the two gaps $E^{op}_{gap}$ and $E^{sp}_{gap}$ differ from each other by less 
than 5\% in the case of SiC substrate.

To see why the optical gap is insensitive to both the long-range and short-range 
Coulomb interactions, we plot the exciton binding energy in Fig. \ref{4x}. It is found 
that $E_X$ increases from about 0.89~eV to 1.42~eV as the substrate changes from SiC 
to $\mbox{SiO}_2$. This range happens to be similar to the previous first-principles 
calculations on graphene nanoribbons \cite{Yang}. It is reminded that $E^{qp}_{gap}$ 
gains about 0.59~eV when $\epsilon_r^*$ decreases 6.4 to 2.5. In the meantime, due to 
the same substrate change, $E_X$ increases by about 0.53~eV. Considering that the 
quasiparticle gap and exciton binding energy contribute to the optical gap in the 
opposite way, i.e., $ E^{op}_{gap} = E^{qp}_{gap} - E_X $, the net change in the 
optical gap is only $0.59 - 0.53 = 0.06$~eV, one order of magnitude smaller than 
either $E^{qp}_{gap}$ or $E_X$. The Coulombic energy $E_X$ consists mainly of a 
polarization contribution while the quasiparticle gap is largely determined by the 
self-energy contribution. Although both terms are shown to depend strongly on the 
dielectric environment, what is most interesting here is that the quasiparticle and 
excitonic effects have very similar dependence on the dielectric constant. In fact, we 
find that the exciton effect is also dominated by the long-range Coulomb interaction. 
The on-site Coulomb interaction gives an exciton binding energy of 0.38~eV, which is 
less than 30\% of the overall exciton binding energy in the case of $\mbox{SiO}_2$ 
substrate.

\begin{figure}
  \includegraphics[width=0.40\textwidth]{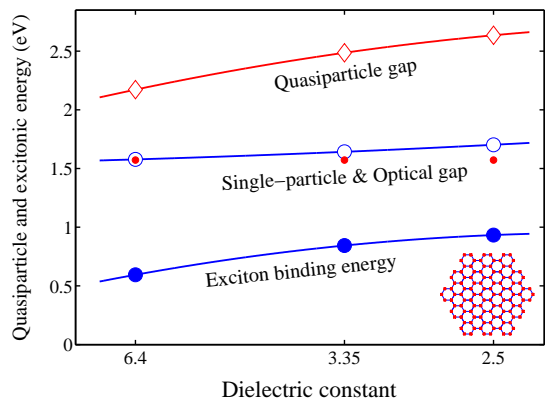}
  \caption{ For a hexagonal nanoflake with $N = 3$ as shown in the inset, quasiparticle
(diamonds) and optical gap (open dots) as well as the exciton binding energy (solid dots)
as a function of the inverse of the effective dielectric constant. The single-particle
gap, which does not vary with the dielectric constant, is shown in the smaller solid dots
as a reference. }
  \label{3x}
\end{figure}

Our next model system is a hexagonal graphene nanoflake with armchair edges which has a 
similar single-particle spectrum to the previous triangular model. Figure \ref{3x} plots 
the quasiparticle and optical gaps together with the exciton binding energy as a function 
of the inverse of dielectric constant. Overall, we find that $E^{qp}_{gap}$, 
$E^{op}_{gap}$, and $E_X$ exhibit very similar dependence on $\epsilon_r^*$ to those seen 
for the triangular model. Specifically, as the substrate changes from SiC to 
$\mbox{SiO}_2$, the quasiparticle correction to the energy gap is seen to increase from 
0.6~eV to 1.06~eV by about 0.46~eV while the exciton binding energy from 0.59~eV to 
0.93~eV by 0.34~eV. As a result, the optical gap gains about 0.12~eV due to the reduced 
screening effect. Furthermore, most noticeably is that the optical gap becomes almost 
identical to the single-particle gap in the case of SiC substrate.

\begin{figure}
  \includegraphics[width=0.40\textwidth]{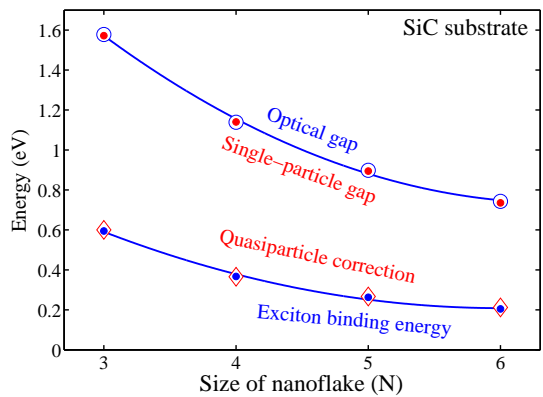}
  \caption{ Single-particle gap (solid dots), quasiparticle correction (diamonds), optical 
gap (open dots) and exciton binding energy (solid dots) calculated as a function of the 
size of the hexagonal nanoflake on a SiC substrate. }
  \label{gap-size}
\end{figure}

Figure \ref{gap-size} shows the size dependence of the single-particle and optical gaps, 
quasiparticle correction to the energy gap and exciton binding energy for the hexagonal 
nanoflake on a SiC substrate. When the size increases, the quantum confinement looses its 
effect and all the gaps as well as the exciton binding energy are seen to gradually 
decrease. As for the quasiparticle effect, let us reformulate Eq. \ref{exciton} as 
follows,
\begin{equation}
E^{op}_{gap} - E^{sp}_{gap} = \left( E^{qp}_{gap} - E^{sp}_{gap} \right) - E_X .
\end{equation}

As the size increases, the quasiparticle correction to the energy gap, i.e., $E^{qp}_{gap} 
- E^{sp}_{gap}$ and the exciton binding energy $E_X$ are found to exhibit nearly the same 
dimensional dependence, which leads to an almost exact cancellation of their contributions 
to the optical gap. As a result, we see that $E^{op}_{gap} - E^{sp}_{gap}$ nearly 
vanishes, i.e., the optical gap closely follows the single-particle gap. The surprising 
overlap of the single-particle and optical gaps can be explained in the following. We have 
seen that a small difference between the quasiparticle correction and exciton binding 
energy is caused by the long-range Coulomb interaction. In the case of SiC substrate where 
the long-range Coulomb interaction is greatly suppressed, both $E^{qp}_{gap} - 
E^{sp}_{gap}$ and $E_X$ are now mainly determined by the short-range Coulomb interaction 
and thus become almost identical to each other.

% \section{Conclusion}

In summary, we have carried out a configuration-interaction study of quasiparticle and 
excitonic effects in graphene nanoflakes on various substrates. We have identified that both 
the quasiparticle correction to the energy gap and exciton binding energy are dominated by 
the long-range Coulomb interactions, and furthermore, these two terms exhibit similar 
dependence on the dielectric constant of the substrate. As a result, their contributions to 
the optical gap almost cancel each other, which leads to a weak dependence of the optical 
gap on the dielectric environment. In the case of substrate with larger dielectric constant 
and thus strong screening effect like SiC, the optical gaps of graphene nanoflakes are 
revealed to closely follow the single-particle gap as if all the electron-electron 
interactions are quenched.

\acknowledgments
This work is supported by National Basic Research Program of China (973 Program No. 2011CB925602).

\end{document}